\documentclass[manuscript]{aastex}

\usepackage{amsbsy,amssymb,amsmath,bm}
\usepackage{graphicx,subfigure}
\usepackage{natbib}
\usepackage{color}
\usepackage{ulem}

\begin{document}

\title{Dynamo induced by time-periodic force}
\author{Xing Wei}
\affil{Institute of Natural Sciences, School of Mathematical Sciences, Shanghai Jiao Tong University}
\email{xing.wei@sjtu.edu.cn}

\begin{abstract}
To understand the dynamo driven by time-dependent flow, e.g. turbulence, we investigate numerically the dynamo induced by time-periodic force in rotating magnetohydrodynamic flow and focus on the effect of force frequency on the dynamo action. It is found that the dynamo action depends on the force frequency. When the force frequency is near resonance the force can drive dynamo but when it is far away from resonance dynamo fails. In the frequency range near resonance to support dynamo, the force frequency at resonance induces a weak magnetic field and magnetic energy increases as the force frequency deviates from the resonant frequency. This is opposite to the intuition that a strong flow at resonance will induce a strong field. It is because magnetic field nonlinearly couples with fluid flow in the self-sustained dynamo and changes the resonance of driving force and inertial wave.
\end{abstract}

\maketitle

\section{Motivation}

Dynamo widely exists in astronomical objects such as planets, stars, galaxies and disks to generate magnetic fields in these objects through the motion of conducting fluids. The driving forces of dynamo are various, e.g. thermal convection, precession, libration, tide, and so on. In the Earth's liquid core or the solar convective zone, or more generally, planetary and stellar interiors, thermal convection drives time-dependent flow which is powerful for the dynamo action. A great amount of reference papers about convection dynamo can be found. Here we list some books and review papers, e.g. \citep{moffatt}, \citep{proctor}, \citep{busse}, \citep{zhang}, \citep{roberts}, \citep{christensen}, \citep{jones}. In addition, the Earth's or planetary precession can drive flow instabilities for the dynamo action \citep{malkus, tilgner, wei-precession, tilgner2016, lin}. Precession and precessional instabilities are both time-periodic. Similarly, planetary libration can also drive the dynamo action \citep{wu}. Libration and librational instabilities are both time-periodic. Tide in a binary system, e.g. two stars or a star and its planet evolving around each other, drives waves in the fluid interior of an astronomical object for the dynamo action \citep{houben, cebron, wei-tide, lin-ogilvie}. Again, tide and tidal waves are both time-periodic. The details about precession, libration and tide can be found in the review papers \citep{ogilvie,lebars}. In summary, convection, precession and precessional instabilities, libration and librational instabilities, tide and tidal waves, are all time-periodic, i.e. these driving forces for dynamo are time-periodic. Moreover, these driving forces have a wide range of time scale. Then how the force frequency influences the dynamo action is an interesting problem for geophysics and astrophysics. In this paper we will study the dynamo induced by time-periodic force in a local periodic box.

On the other hand, all the astronomical objects are rotating, and the Coriolis force due to rotation induces inertial waves which can resonate with the time-periodic force. When the force frequency is close to the frequency of inertial waves, flow becomes very strong. It may be concluded that a strong flow near resonance will induce a strong magnetic field through the dynamo action. However, in this paper we will verify that this conclusion is not completely correct.

In \S2 the model is built, in \S3 the numerical results are shown and discussed, and in \S4 a brief summary is given.

\section{Model}

We consider an incompressible and conducting fluid rotating with angular velocity $\bm\Omega$ in a periodic box. The periodic box can be considered as a small piece taken from the interior of an astronomical object, i.e. the so-called local box. The fluid flow is driven by a time-periodic force $\bm f$. The frequency of driving force, which represents the frequency of convective rolls or inertial waves in precessional or tidal flow, is too slow compared to that of sound wave. Moreover, although compressibility changes the pattern of convection, it does not play an important role in the dynamo action because sound wave is much faster than the time scale of magnetic field variation. Therefore, we consider an incompressible fluid. The dimensionless equation of fluid motion in rotating frame reads
\begin{equation}\label{eq:ns}
\frac{\partial\bm u}{\partial t}+\bm u\cdot\bm\nabla\bm u=-\bm\nabla p+E\nabla^2\bm u+2\bm u\times\hat{\bm z}+(\bm\nabla\times\bm B)\times\bm B+\bm f.
\end{equation}
In Equation \eqref{eq:ns}, the two terms on the left-hand-side are the local and advective accelerations of fluid motion. The terms on the right-hand-side are successively the pressure gradient incorporating the centrifugal potential, the viscous force, the Coriolis force due to rotation, the Lorentz force due to magnetic field, and the driving force. Length is normalized with $l'=l/(2\pi)$, i.e. the box size $l$ divided by $2\pi$, such that the dimensionless box size is $2\pi$, time with the inverse of rotational frequency $\Omega^{-1}$, velocity with $\Omega l'$, pressure with $\rho(\Omega l')^2$ ($\rho$ being fluid density), and magnetic field with $\sqrt{\rho\mu}\Omega l'$ ($\mu$ being magnetic permeability). The dimensionless equation of magnetic field reads
\begin{equation}\label{eq:induction}
\frac{\partial\bm B}{\partial t}=\bm\nabla\times(\bm u\times\bm B)+\frac{E}{Pm}\nabla^2\bm B.
\end{equation}
In Equation \eqref{eq:induction} the two terms on the right-hand-side are the induction and diffusion terms. The two dimensionless numbers are the Ekman number $E=\nu/(\Omega l'^2)$ ($\nu$ being viscosity) which is the ratio of the rotational time-scale $\Omega^{-1}$ to the viscous time-scale $l'^2/\nu$ and measures the strength of rotation, and the magnetic Prandtl number $Pm=\nu/\eta$ which is the ratio of viscosity $\nu$ to magnetic diffusivity $\eta$.

The driving force $\bm f$ in Equation \eqref{eq:ns} is modeled as a single plane wave \citep{wei-resonance}, 
\begin{equation}
\bm f=\Re\{\hat{\bm f}\exp{i(\bm k\cdot\bm x-\omega t)}\},
\end{equation} 
where $\hat{\bm f}$ is the complex force amplitude, $\bm k$ the force wavevector which should be an integer to keep the periodic boundary condition, $\omega$ the force frequency, and $\Re$ denotes the real part. To have the dynamical effect, the driving force should be vortical, and moreover, we assume it a helical force, i.e. $\bm\nabla\times\bm f=k\bm f$, which tends to drive a helical flow in the linear regime and almost a helical flow in the weakly nonlinear regime. Although the helical flow is not necessary for the dynamo action it facilitates the dynamo action \citep{wei2014gafd}. This assumption of helical force is reasonable because any force field can be decomposed to a potential part absorbed into the pressure gradient and a solenoidal part expressed as the superposition of helical forces \citep{waleffe}. It should be noted that $\hat{\bm f}$ is a complex amplitude which contains the phase information. The details about the driving force can be found in \citet{wei-resonance}.

In \citet{wei2014gafd} it was numerically proved that helicity is, though not necessary, of help to dynamo action, and a small-scale helical flow driven by a high force wavenumber $k$ tends to induce a large-scale magnetic field. In our numerical calculations with the driving force on $k=1$, the magnetic energy concentrates on large scales and the highest magnetic energy is on the magnetic wavenumber $k_m=1$. Certainly, a larger $k$ is better for dynamo action, and in this paper we do not investigate the length scale but the time scale of driving force. It should be noted that our self-sustained dynamo is different from MHD turbulence in which a large-scale magnetic field is externally imposed. In MHD turbulence the small scales tend to be suppressed due to the anisotropy arising from the propagation of Alfv\'en waves \citep{haugen, wei2016acta}.

The outputs are the dimensionless kinetic and magnetic energies, both of which are normalized with $\rho\Omega^2l'^5$. We use the spectral method for the numerical calculations. For numerical convergence the sufficient resolutions are guaranteed that both the kinetic and magnetic energy spectra decay by more than ten magnitudes, namely the highest energy contained in the low mode (not necessarily $k_x=k_y=k_z=1$ but a small wavenubmer) is more than $10^{10}$ times of the lowest energy contained in the highest mode. In the regime of the moderate parameters we study (see the next section), the resolution $128^3$ is used. In this paper we do not pursue the scale of numerical calculations but try to reveal the physics by our moderate-scale numerical calculations.

\section{Results}

We focus on the effect of force frequency on the dynamo action, and thus we vary the force frequency $\omega$ but fix all the other parameters. The Ekman number $E$ is fixed to be $10^{-3}$. The typical value of $E$ based on the global length scale, e.g. the planetary or stellar radius, is much smaller than $10^{-3}$. But what we consider is a local box with its size $l$ much smaller than the global length scale, and the local $E$ at the order $10^{-3}$ is already sufficiently small to represent the rapid rotation of an astronomical object. $Pm$ is fixed to be unity for the numerical reason (too high or too low $Pm$ will bring numerical stiffness). The force amplitude $|\hat{\bm f}|$ is fixed to be $5\times 10^{-3}$ which is sufficiently high to drive a dynamo. A stronger force amplitude will support a dynamo with higher magnetic energy but in this paper we focus on the effect of force frequency on dynamo and do not test more force amplitudes. The force wave vector is fixed to be ($k_x=k_y=k_z=1$) which stands for a small length scale $l/\sqrt{3}$.

In a rapidly rotating fluid, the inertial wave frequency is $\pm 2\bm\Omega\cdot\bm k/k$. In our model the resonance between the driving force and inertial wave occurs at the frequency $\omega_0=-2\bm\Omega\cdot\bm k/k$ and its dimensionless value is $-2k_z/k=-2/\sqrt{3}\approx-1.1547$. The details about the resonant frequency can be found in \citep{wei-tide, wei-resonance}. We investigate the force frequencies around $\omega_0$ until dynamo fails. When $\omega\ge-1.13$ and $\omega\le-1.19$, dynamo fails. It is reasonable because the force frequency too far away from resonance induces too weak flow to induce dynamo. Next, we study the frequency range near resonance which supports the dynamo action.

\begin{figure}
\centering
\includegraphics[scale=0.9]{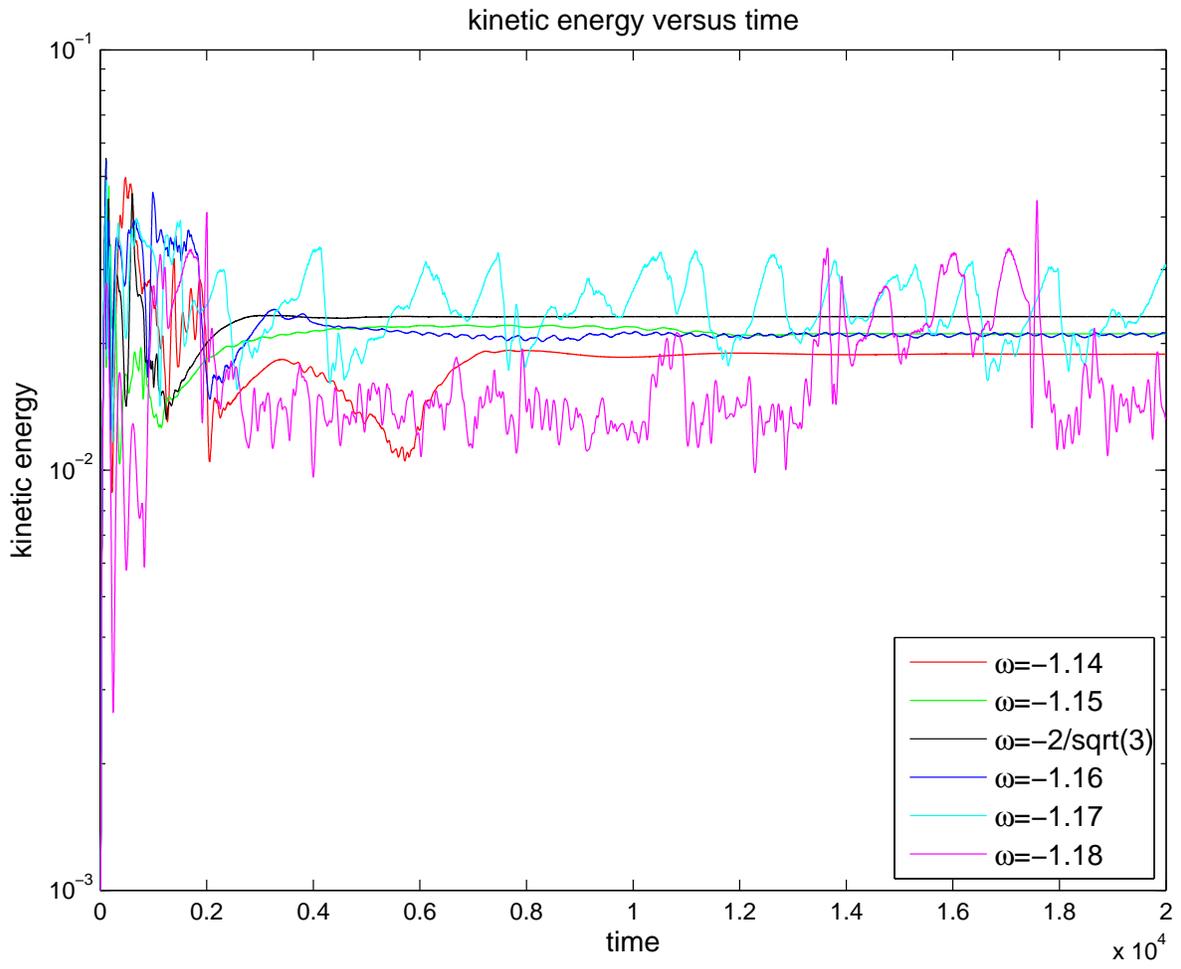}
\caption{Kinetic energy versus time at different force frequencies.}\label{fig:ke}
\end{figure}

\begin{figure}
\centering
\includegraphics[scale=0.9]{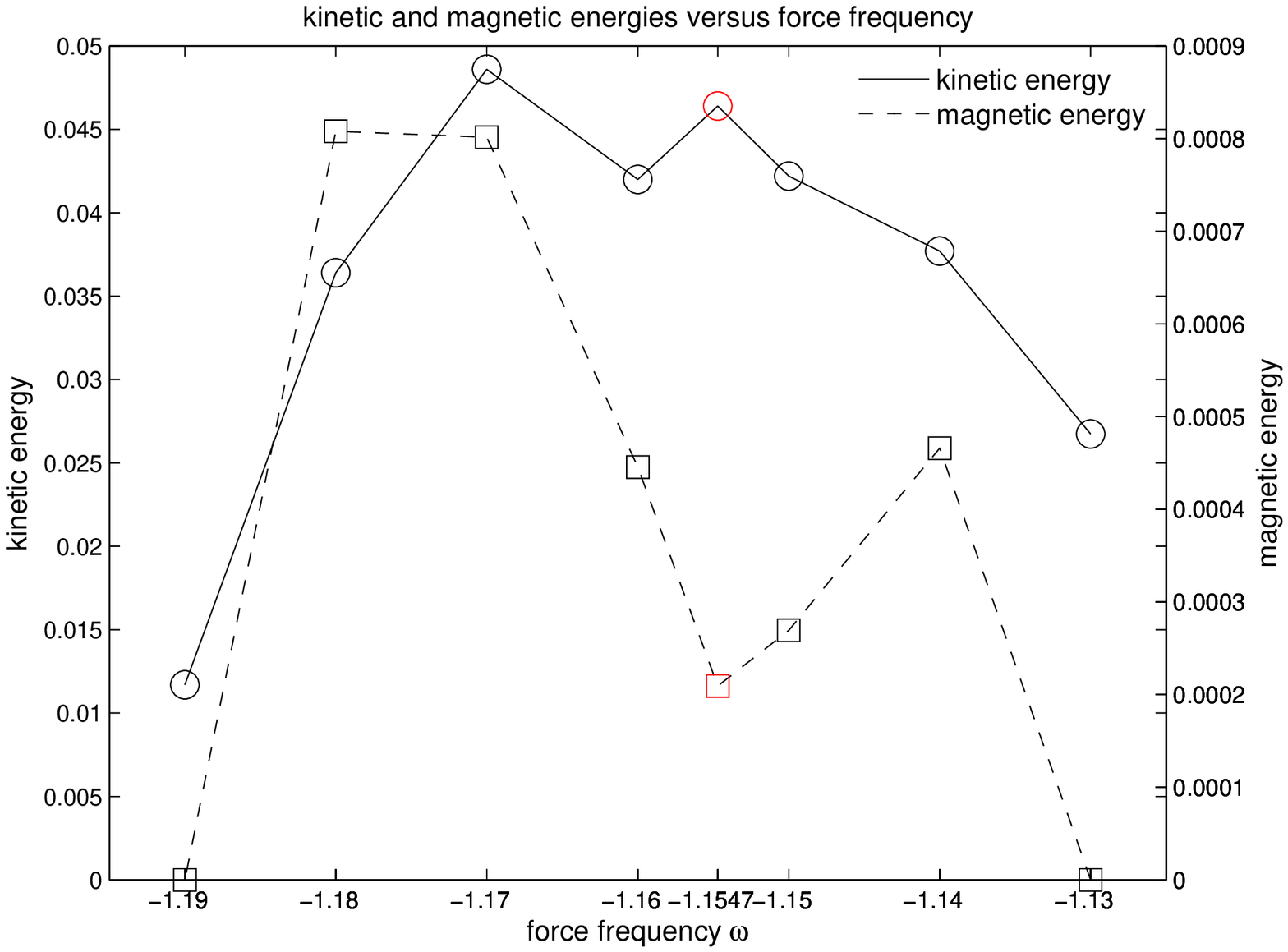}
\caption{Mean kinetic and magnetic energies versus force frequency. Solid line denotes kinetic energy and dashed line for magnetic energy. Left vertical axis shows the value of kinetic energy and right axis for magnetic energy. The red symbols denote the energies at the frequency of inertial wave $\omega_0=-2/\sqrt{3}\approx -1.1547$.}\label{fig:energy}
\end{figure}

We numerically calculate dynamo equations until time=$2\times 10^4$, which is 20 magnetic diffusion time scales and sufficiently long for a self-sustained dynamo to saturate. Figure \ref{fig:ke} shows kinetic energy versus time at different force frequencies from -1.14 to -1.18 where $\omega_0$ lies in between. When the force frequency is out of this range, the dynamo fails. At $\omega_0\approx-1.1547$ and the two lower frequencies -1.15 and -1.14, when kinetic energy saturates it does not significantly vary with time. However, at the higher frequencies it varies with time. At $\omega=-1.16$ it oscillates but the oscillation amplitude is small. At $\omega=-1.17$ and -1.18 the oscillation amplitudes are large and can reach 30\% of the mean values. The mean kinetic energy versus frequency is plotted with the solid line in Figure \ref{fig:energy}. It is shown that the kinetic energy at $\omega_0$ is the second largest and except $\omega=-1.17$ the kinetic energies decrease as the frequency distance $|\Delta\omega|=|\omega-\omega_0|$ increases.

\begin{figure}
\centering
\includegraphics[scale=0.9]{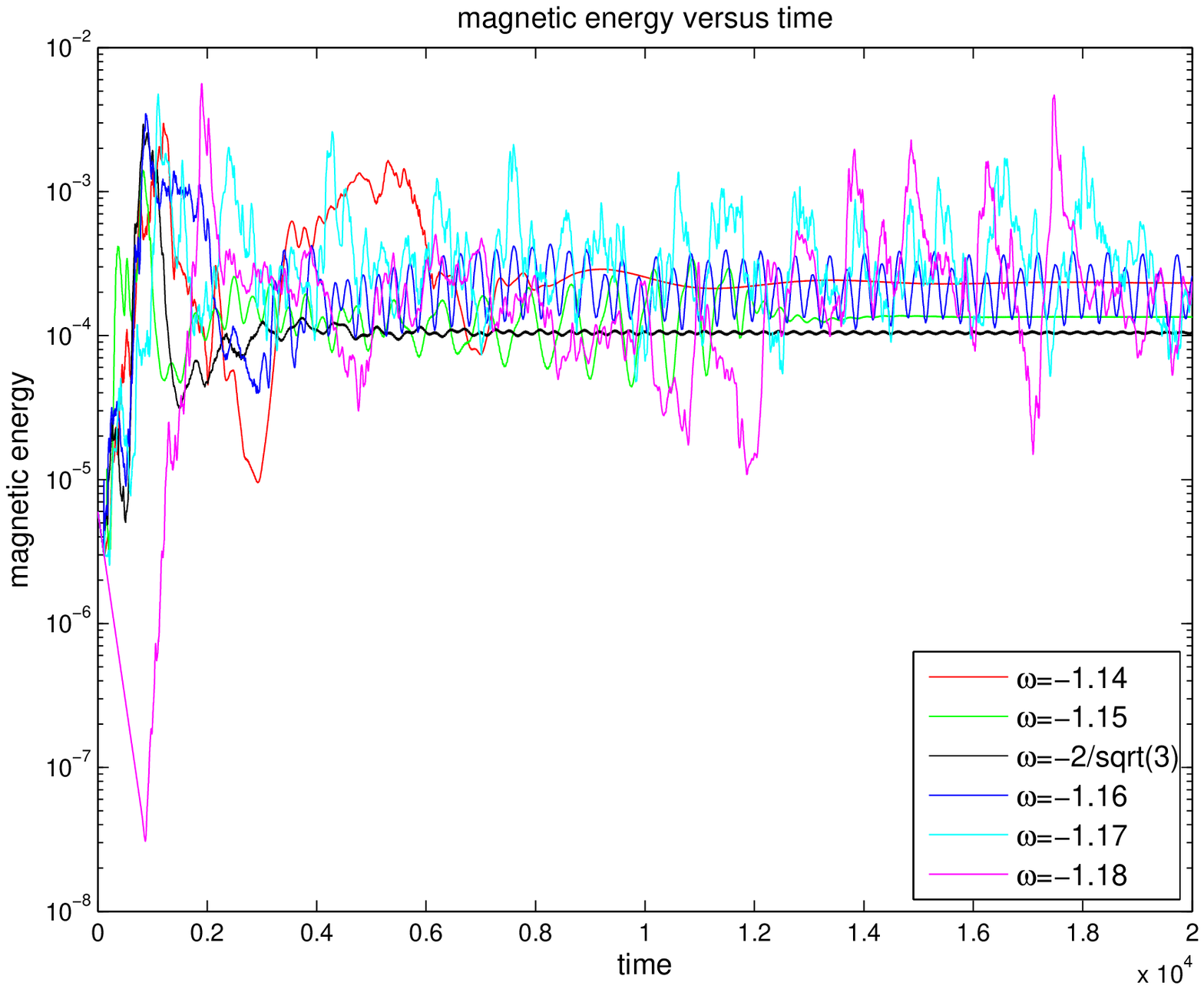}
\caption{Magnetic energy versus time at different force frequencies.}\label{fig:me}
\end{figure}

Figure \ref{fig:me} shows magnetic energy versus time at different force frequencies. At $\omega_0$ and the two lower frequencies -1.15 and -1.14 magnetic energy does not significantly vary with time but varies greatly at the higher frequencies -1.16, -1.17 and -1.18, and a higher frequency leads to higher fluctuations. It is interesting that the mean magnetic energy is minimum at $\omega_0$ and increases as the frequency distance $|\Delta\omega|$ increases, as plotted with the dashed line in Figure \ref{fig:energy}.

We now move to Figure \ref{fig:energy} which has already been discussed in the last two paragraphs. The mean energies are obtained by taking the average when the energies have already saturated. The red symbols show the mean energies at $\omega_0$. This figure clearly indicates that at the resonant frequency $\omega_0$ magnetic energy reaches its minimum value although kinetic energy is strong. Moreover, the force frequency more distant away from $\omega_0$ leads to a stronger magnetic energy. At $\omega=-1.14$ and -1.18, the two most distant frequencies away from $\omega_0$, magnetic energy is strong although kinetic energy is weak. This result is opposite to the intuition that a stronger flow will induce a stronger magnetic field. 

The reason is unclear but we may give a tentative explanation. It is probably because magnetic field changes the resonance. In the linear regime the presence of magnetic field changes the resonant frequency \citep{wei-tide}, and in the nonlinear regime magnetic field can change the saturation level of magnetic energy in the dynamo action via the nonlinear coupling of flow and field, i.e. the Lorentz force in Equation \eqref{eq:ns} and the induction term in Equation \eqref{eq:induction}. This nonlinear suppression effect near resonance has been already found in the hydrodynamic case \citep{wei-resonance}.

\begin{table}
\centering
\begin{tabular}{|c|c|c|c|c|c|c|}
\hline
$\omega$ & -1.14 & -1.15 & $-2/\sqrt{3}$ & -1.16 & -1.17 & -1.18 \\
\hline
$Re=Rm$ & 110 & 116 & 122 & 116 & 124 & 108 \\
\hline
\end{tabular}
\caption{$Re=Rm$ for different force frequencies.}\label{tab:Re}
\end{table}

We then investigate Reynolds number $Re$ and magnetic Reynolds number $Rm$. We define the characteristic velocity $U$ as the square root of volume-averaged $u^2$ and then $Re=Ul/\nu=2\pi\sqrt{\int u^2dV/V}/E$ (remember that $l=2\pi l'$). In our calculations $Pm=1$ so that $Rm=RePm=Re$. Table \ref{tab:Re} lists $Re=Rm$ for different force frequencies. The difference of $Re=Rm$ values for different force frequencies is not very large, and the values are higher than 100, which is sufficient for turbulent flow (in terms of $Re$) and for dynamo action (in terms of $Rm$) in a periodic box.

\begin{figure}
\centering
\subfigure[]{\includegraphics[scale=0.8]{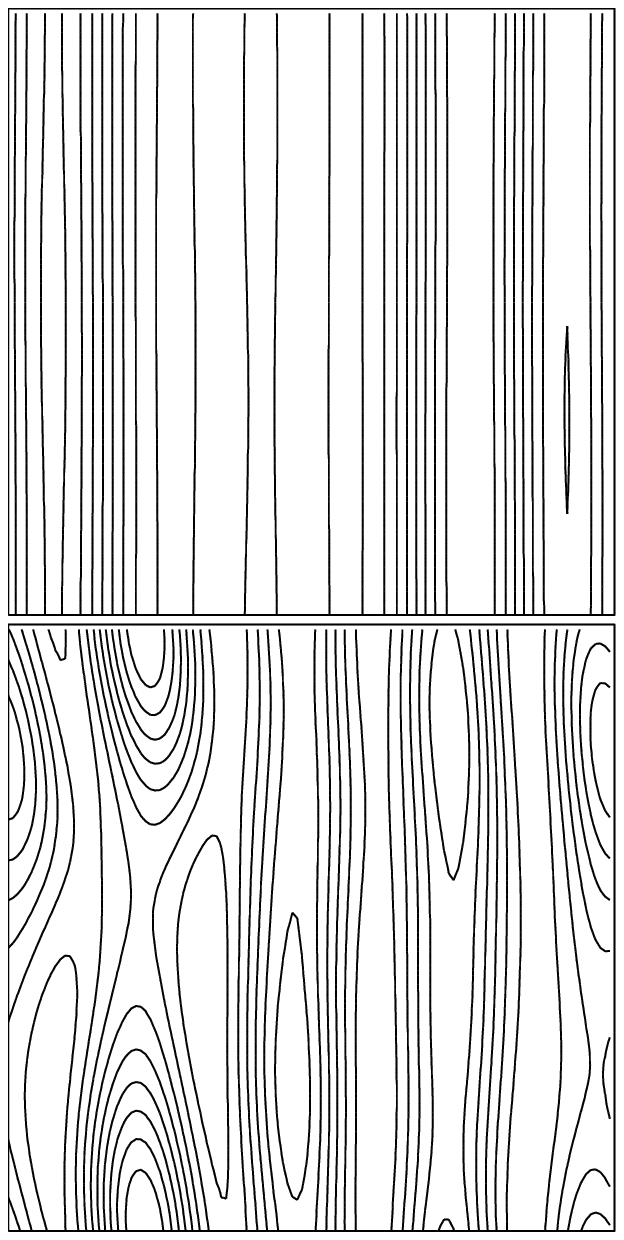}\label{fig:xz1}}
\subfigure[]{\includegraphics[scale=0.8]{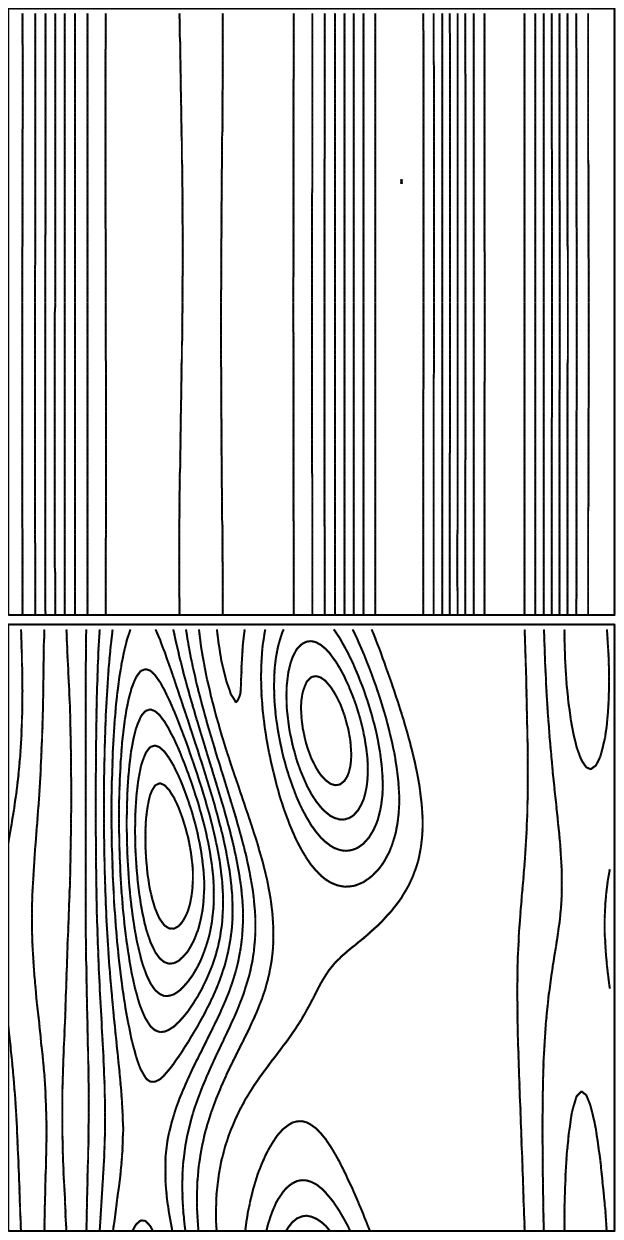}\label{fig:xz2}}
\subfigure[]{\includegraphics[scale=0.8]{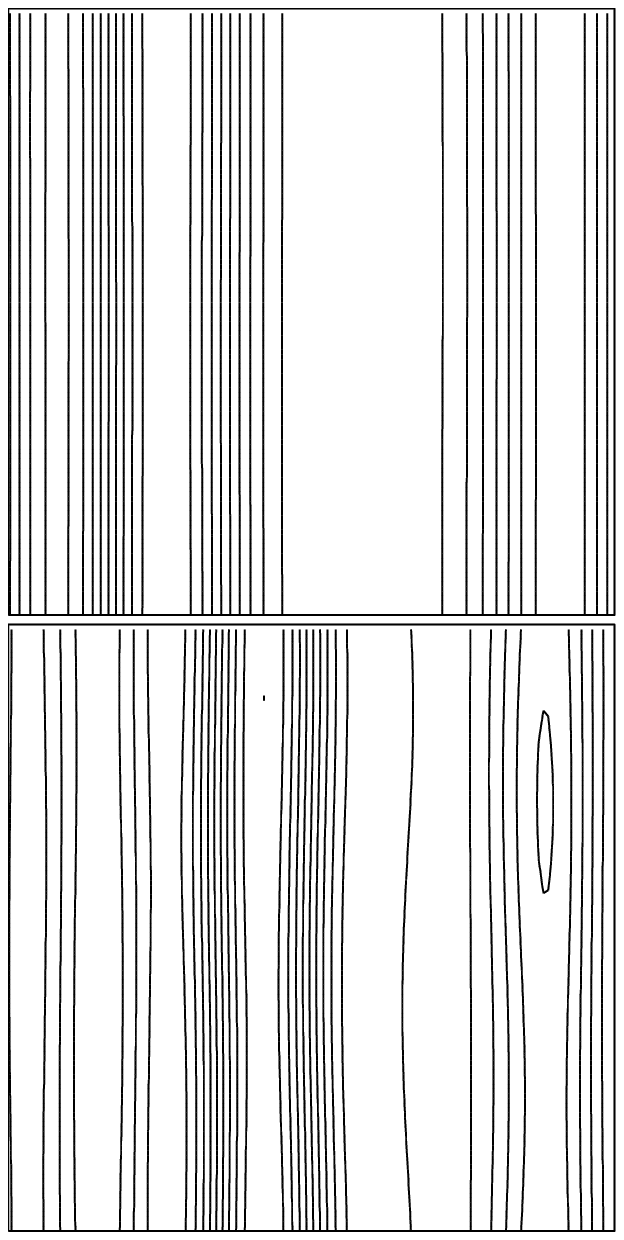}\label{fig:xz3}}
\caption{Contours of kinetic energy density (top) and magnetic energy density (bottom) in x-z plane at $y=0$. (a) $\omega=-1.14$, (b) $\omega=-2/\sqrt{3}$, (c) $\omega=-1.18$. Time=$2\times 10^4$.}\label{fig:xz}
\end{figure}

Next we investigate the distributions of velocity and magnetic field. Figure \ref{fig:xz} shows the contours of kinetic energy density and magnetic energy density in the x-z plane at $y=0$ at three force frequencies, $-1.14$, $-2/\sqrt{3}$ and $-1.18$, and at the saturation level (time=$2\times 10^4$). The distribution of velocity follows the Taylor-Proudman theorem (z independent and columnar structure) at all the three frequencies because of the rapid rotation. However, the distributions of magnetic field are quite different at the three force frequencies. At the resonant frequency (Fig. \ref{fig:xz2}) the distribution of magnetic field deviates from the columnar structure whereas it tends to have the columnar structure at the frequencies away from resonance (Figs. \ref{fig:xz1} and \ref{fig:xz3}). This implies that magnetic field at resonance leads to a strong field distortion such that the strong Ohmic dissipation suppresses the dynamo efficiency.

\begin{table}
\centering
\begin{tabular}{|c|c|c|c|c|c|c|}
\hline
$\omega$ & -1.14 & -1.15 & $-2/\sqrt{3}$ & -1.16 & -1.17 & -1.18 \\
\hline
growth rate & 0.0301 & 0.0268 & 0.0262 & 0.0247 & 0.0218 & 0.0184 \\
\hline
\end{tabular}
\caption{Growth rate of kinematic dynamo for different force frequencies.}\label{tab:growth}
\end{table}

\begin{figure}
\centering
\includegraphics[scale=0.9]{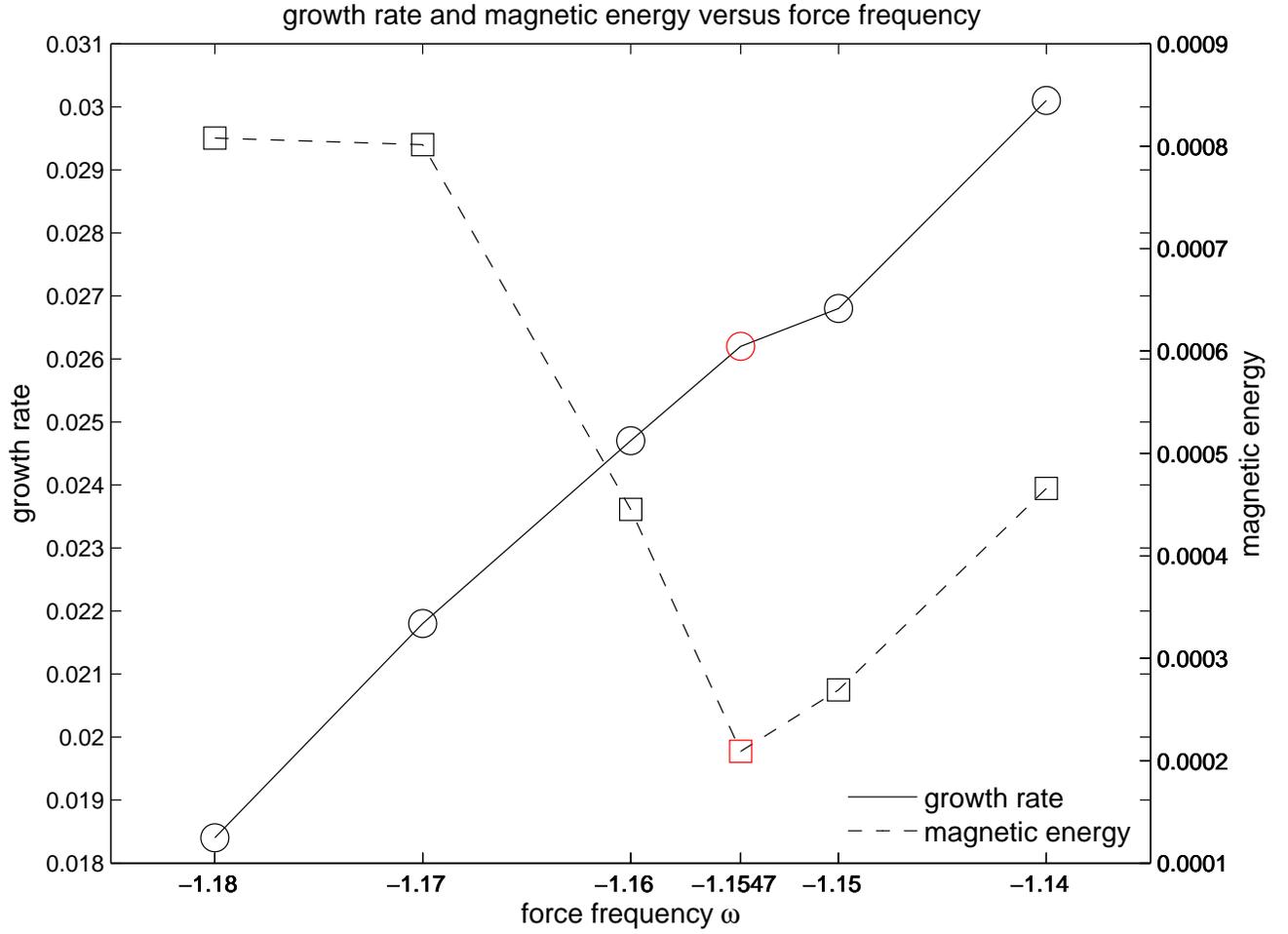}
\caption{As in Figure \ref{fig:energy} except that the solid line and the left axis denote the growth rate of kinematic dynamo.}\label{fig:growth}
\end{figure}

\begin{figure}
\centering
\includegraphics[scale=0.9]{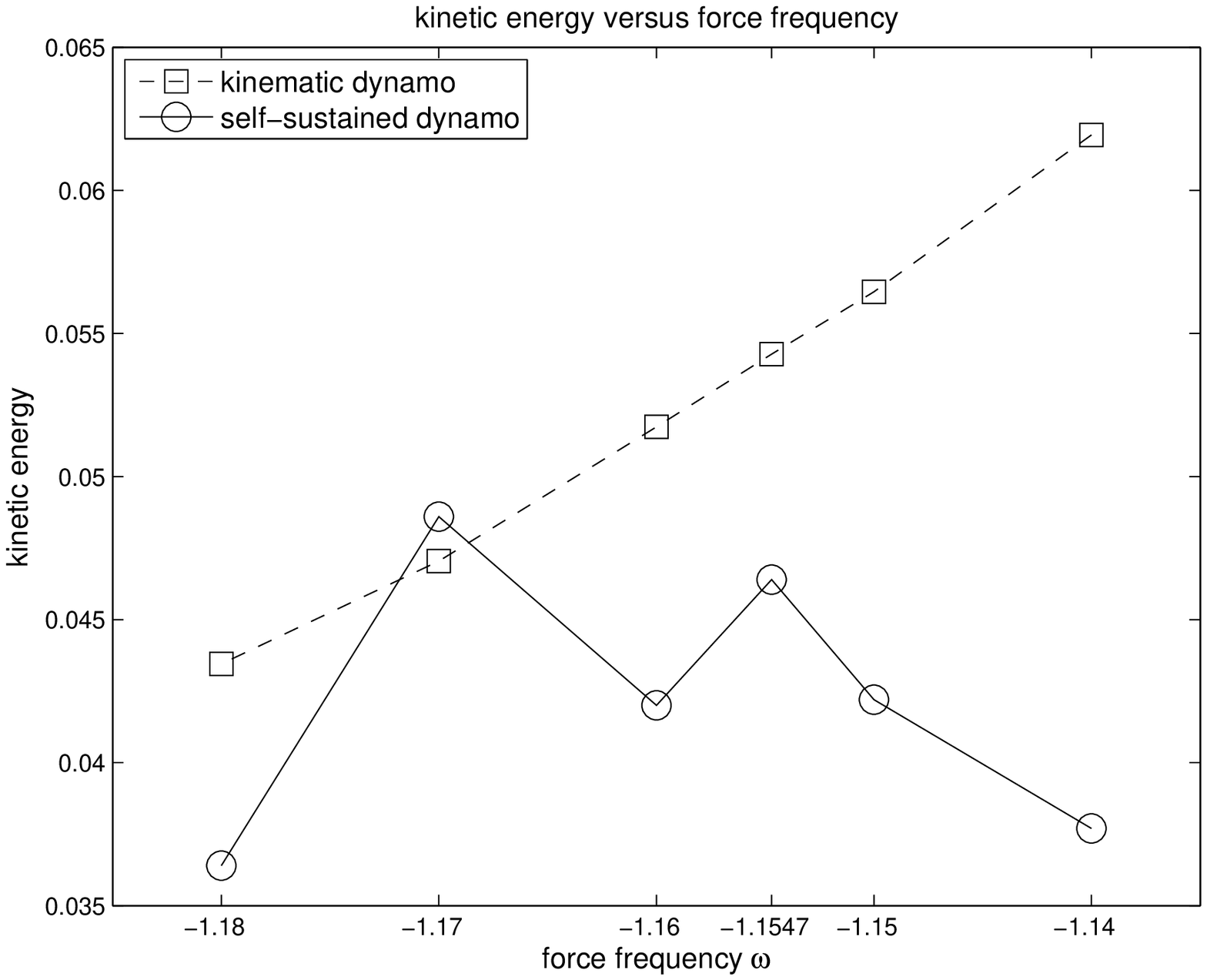}
\caption{Kinetic energy versus force frequency for kinematic dynamo and self-sustained dynamo.}\label{fig:kinetic}
\end{figure}

To better understand the self-sustained dynamo, we go on calculating the kinematic dynamo for comparison. We simply drop the Lorentz force (the second last term) in Equation \eqref{eq:ns}. Table \ref{tab:growth} shows the growth rate of kinematic dynamo for different force frequencies. It indicates that the growth rate of kinematic dynamo decreases with the absolute value of frequency $|\omega|$ increasing. Figure \ref{fig:growth} shows the comparison between kinematic dynamo and self-sustained dynamo. The solid line denotes growth rate of kinematic dynamo while the dashed line denotes magnetic energy of self-sustained dynamo. Clearly, self-sustained dynamo behaves in a different manner from kinematic dynamo. This implies that the lowest magnetic energy at resonance could be because magnetic field changes the dispersion relation of wave (with the neglect of Lorentz force in the kinematic dynamo, the dispersion relation of wave cannot be changed). As another evidence, Figure \ref{fig:kinetic} shows kinetic energy versus force frequency for kinematic dynamo and self-sustained dynamo. Again, the kinematic dynamo and self-sustained dynamo behave quite differently, namely the Lorentz force changes the resonance of inertial wave and time-periodic force.

\section{Summary}

In this paper we numerically solve rotating magnetohydrodynamics equations in a local box to investigate the effect of force frequency on the dynamo action to understand the dynamo driven by time-dependent flow. The force frequency far away from resonance cannot support the dynamo action. More interestingly, it is opposite to the intuition that a strong flow at resonance will induce a strong magnetic field, instead, a strong flow at resonance induces a weak field. Moreover, within the dynamo range near resonance, the force frequency more distant from the resonant frequency leads to a stronger magnetic energy. The reason is that the presence of magnetic field changes the resonance and that the field distortion at resonance suppresses the dynamo efficiency.

\section*{Acknowledgements}
An anonymous referee gave me good comments and suggestions. The work was supported by the grant of 1000 youth talents of the Chinese government.

%\bibliographystyle{apj}
%\bibliography{paper}

\end{document}